%% file: main.tex
\documentclass[10pt, conference]{IEEEtran}
\IEEEoverridecommandlockouts
\usepackage{cite}
\usepackage{amsmath,amssymb,amsfonts}
\usepackage{algorithmic}
\usepackage{graphicx}
\usepackage{textcomp}
\usepackage{tablefootnote}
\usepackage{booktabs}
\usepackage{url}
\usepackage{bm}
\usepackage{color}
\usepackage{multirow}
\usepackage{subfigure}
\usepackage{xcolor}
\usepackage[
]{hyperref}
\newcommand{\figureref}[1]{Fig.~\ref{#1}}

\newcommand{\secref}[1]{Sec.~\ref{#1}}

\newcommand{\red}[1]{\textcolor{black}{#1}}
\makeatletter
\newcommand{\figcaption}[1]{\def\@captype{figure}\caption{#1}}
\newcommand{\tblcaption}[1]{\def\@captype{table}\caption{#1}}

\def\BibTeX{{\rm B\kern-.05em{\sc i\kern-.025em b}\kern-.08em
    T\kern-.1667em\lower.7ex\hbox{E}\kern-.125emX}}
\begin{document}

\title{CloudEmu: A Trace-Driven Cloud-Native Emulation Testbed for Vehicle Video Uplink over Cellular Networks \vspace{-2mm}}

\author{%
  \IEEEauthorblockN{%
    Takashi Torii\textsuperscript{\dag}, 
    Soto Anno\textsuperscript{\dag},
    Masaki Okada, 
    Takuma Tsubaki,
    Nobuhiro Azuma, and Takuya Tojo
  }
  \IEEEauthorblockA{
    {\it Network Service Systems Labs., NTT, Inc.}, Tokyo, Japan \\
    E-mail: \{takashi.torii, soto.anno, masaki.okada, takuma.tsubaki, nobuhiro.azuma, takuya.tojo\}@ntt.com
  }
  \vspace{-7mm}
}

\maketitle

\begingroup
\renewcommand\thefootnote{\dag}%
\footnotetext{S. Anno equally contributed to this work as T. Torii.}%
\endgroup

\input{components/00abstract-en.tex}


\input{components/01intro.tex}

\input{components/02proposed.tex}

\input{components/03demo.tex}

\bibliographystyle{IEEEtran}
\bibliography{references}

\end{document}

%% file: components/00abstract-en.tex
\begin{abstract}
  We present CloudEmu, a trace-driven, cloud-native cellular-emulation testbed for vehicle video uplink communication. Reliable, low-latency video uplink over cellular networks is essential for remote monitoring of autonomous vehicles. However, existing testbeds fall into two extremes. Physical-vehicle platforms provide realism but are costly and make validation under identical network conditions difficult, whereas simulations are inexpensive and reproducible but generally cannot replay field-measured end-to-end performance dynamics without substantial calibration or readily run production video-uplink stacks. A software-defined, cloud-native emulation approach can combine the fidelity of trace-driven replay with the agility and scalability that network softwarization principles offer. To this end, we propose CloudEmu that replays time-synchronized cellular and position traces, collected once from vehicles, on commodity Linux-based virtual vehicle and video-receiver nodes. A Linux-based emulation framework couples traffic replay with position replay, tying network dynamics to each point along the route and enabling repeatable, route-aware experiments without repeated on-road trials. Our demo deploys a production-grade video-uplink stack on CloudEmu, allowing attendees to experience low-cost, repeatable trials and controlled comparisons under identical replayed network conditions.
\end{abstract}

%% file: components/01intro.tex
\section{Introduction}
\label{sec:intro}

As the deployment of autonomous vehicles progresses, ensuring their safe operation increasingly relies on remote monitoring. In this system, a remote operator monitors vehicles through continuous uplink video streams and may intervene when needed~\cite{amador2022survey}. Thus, remote monitoring of autonomous vehicles requires low-latency and highly reliable communication over commercial cellular networks on public roads. These requirements have motivated both academia and industry to develop testbeds for evaluating vehicle video-uplink communication stacks for remote monitoring~\cite{amador2022survey}.

Most existing testbeds fall into two categories: physical-vehicle platforms and simulation-based environments~\cite{charpentier2022assessing,zhang2020v2xsim,cislaghi2023simulation}. The former often use real vehicles on public roads~\cite{charpentier2022assessing}, incurring high operational costs for repeated trials and providing limited reproducibility under dynamic cellular conditions. Researchers have also explored scaled-down vehicles on dedicated tracks to reduce testing costs~\cite{elmoghazy2024real}. However, the reduced size limits the driving area and speed, hindering the evaluation of mobility-induced effects such as cellular handovers that can significantly affect uplink video quality.

Simulation-based testbeds model vehicles, wireless channels, and applications entirely in software to avoid the cost of on-road trials~\cite{zhang2020v2xsim,cislaghi2023simulation}. However, they often either (i) require substantial model tuning and validation against field measurements to match the time-varying, end-to-end performance dynamics of commercial cellular networks, or (ii) provide only limited support for running production video-uplink stacks.

To bridge the gap between costly on-road trials and the limitations of simulators, we propose CloudEmu, a trace-driven testbed for replaying vehicle video uplink over cellular networks. CloudEmu runs virtual vehicle and video-receiver nodes on commodity Linux-based virtual machines (VMs) and connects them through Linux-based network emulator nodes. This emulator replays field-measured, user-plane end-to-end traces of network performance, such as throughput and estimated one-way delay. 
To achieve realistic network emulation, we further introduce a delay correction method that avoids overestimating the emulated delay. CloudEmu also replays position traces, aligning them by timestamps and tying network dynamics to each position along the route.
By capturing cellular and position traces from physical vehicles once per route or test scenario, CloudEmu can repeatedly replay the recorded traces without further on-road driving.

CloudEmu offers three advantages. 
(1) \textbf{Programmable emulation}: CloudEmu reduces testing costs by replaying field-measured, trace-driven user-plane end-to-end network dynamics in a software-defined virtual environment. (2) \textbf{Reproducibility}: Once the traces have been measured, CloudEmu can replay identical conditions, enabling repeatable testing that decouples experimentation from the variability of live cellular deployments. (3) \textbf{Scalability}: CloudEmu is deployable on commodity Linux-based VMs provisioned through infrastructure-as-code, allowing us to scale the number of emulated vehicles and run many experiments in parallel.

Our demo deploys the video-uplink stack used in our SAE Level-4 autonomous-driving trials on CloudEmu, including an adaptive video bitrate (ABR) controller~\cite{okada2025video} and multipath transport~\cite{tojo2025multipath}, to emulate streaming from in-vehicle cameras to a remote video receiver. Attendees can adjust video-delivery parameters and compare the resulting video quality and latency under identical replayed trace conditions, mirroring a tuning process that previously required costly on-road trials.

%% file: components/02proposed.tex
\begin{figure*}[t]
    \centering
    \includegraphics[width=0.8\linewidth]{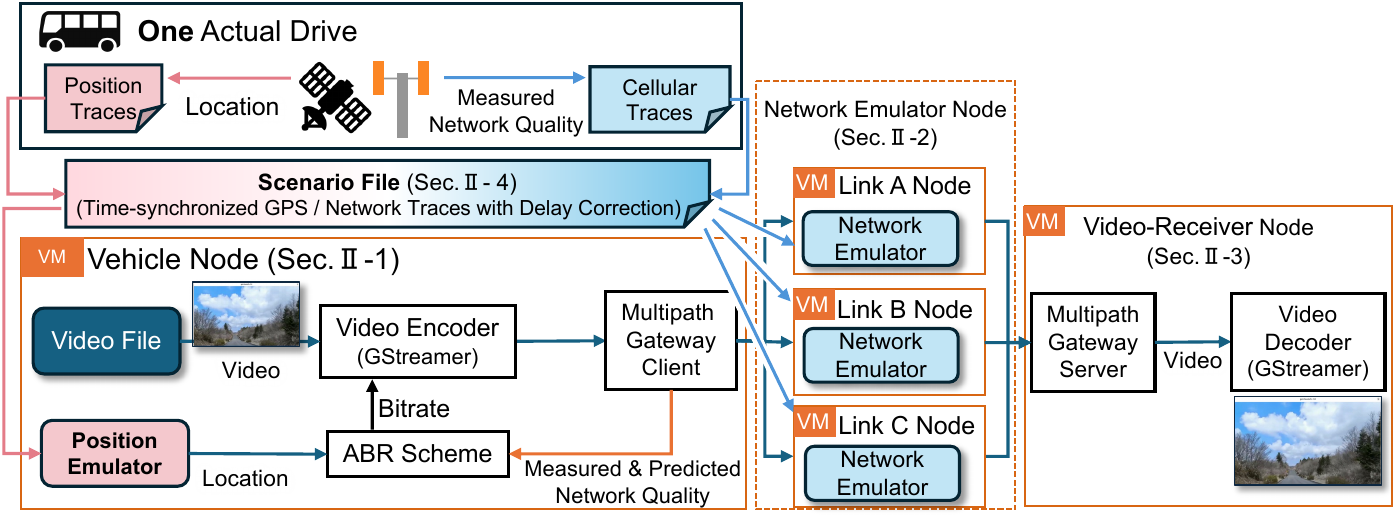}
    \vspace{-4mm}
    \caption{Overview of CloudEmu architecture.}
    \label{fig:cloudemu-overview}
    \vspace{-4mm}
\end{figure*}

\section{Proposed System: CloudEmu}
\label{sec:cloudemu}

As illustrated in \figureref{fig:cloudemu-overview}, CloudEmu consists of three node types: Vehicle Node (\secref{sec:vehicle-node}), Network Emulator Node (\secref{sec:emulator-node}), and Video-Receiver Node (\secref{sec:receiver-node}). These nodes operate based on an emulation scenario derived from GNSS position and network performance data collected during a real drive (detailed in \secref{sec:scenario-file}).
We deploy all nodes as Linux-based virtual machines, provisioned using Terraform\footnote{\url{https://developer.hashicorp.com/terraform}} in an infrastructure-as-code style. This lets us spin up clean environments, tear them down afterward to reduce costs, and run multiple identical setups in parallel.

\subsection{Three Node Types in CloudEmu}

\subsubsection{\textbf{Vehicle Node}}
\label{sec:vehicle-node}

\begin{figure*}[t]
    \begin{center}
        \subfigure[Vehicle-side.]
    {
        \label{fig:gui-vehicle-side}
        \begin{minipage}{0.43\hsize}
            \centering          
            \includegraphics[width=\linewidth]{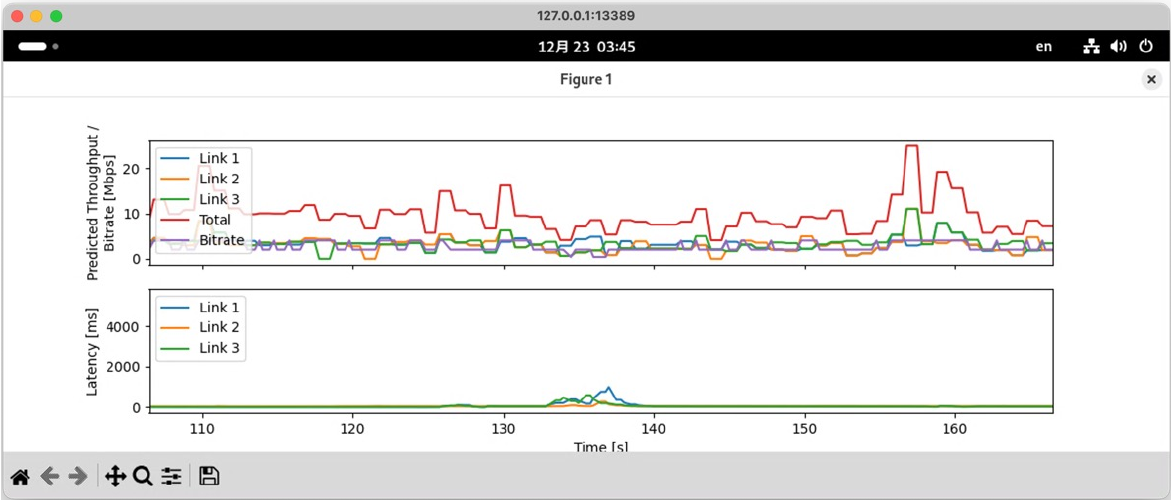}
        \end{minipage}
    }
    \subfigure[Video-receiver-side.]
    {
        \label{fig:gui-center-side}
        \begin{minipage}{0.43\hsize}
            \centering          
            \includegraphics[width=\linewidth]{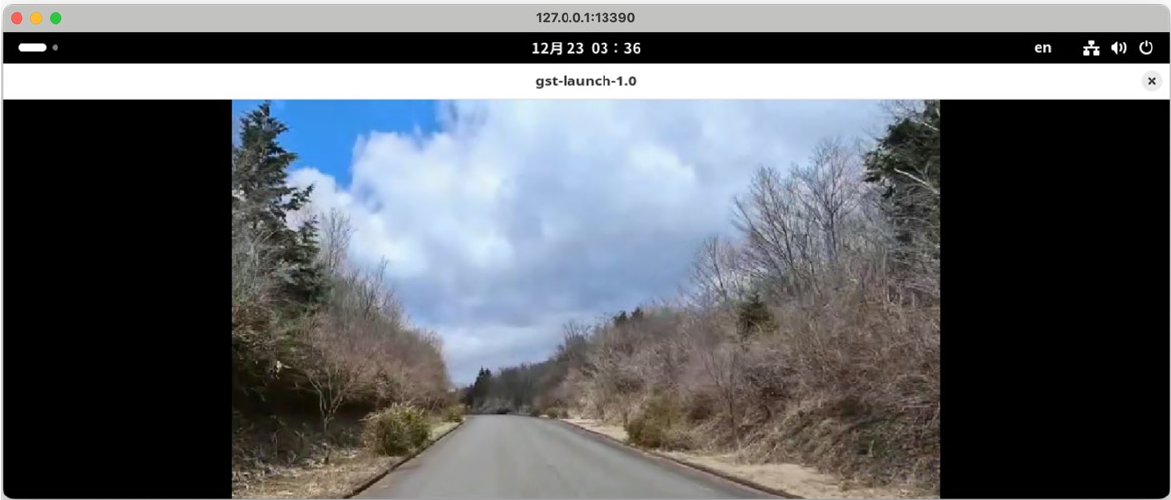}
        \end{minipage}
    } 
    \end{center}
    \vspace{-4mm}
    \caption{Example screens on a vehicle-side and \red{video-receiver-side.}}
    \label{fig:guis}
    \vspace{-4mm}
\end{figure*}

This node emulates a vehicle and its onboard video-uplink stack, including a multipath gateway client, an ABR-based video pipeline, and a position emulator. The position emulator publishes GNSS-like coordinates, synchronized with the emulated network conditions, to the Video-Receiver Node. As shown in \figureref{fig:gui-vehicle-side}, the vehicle-side GUI visualizes ABR behavior by plotting per-path throughput, delay, and the selected video bitrate over time, enabling developers to inspect how bitrate decisions track network changes along the route.

\subsubsection{\textbf{Network Emulator Node}}
\label{sec:emulator-node}

This node represents a cellular path between the Vehicle Node and the Video-Receiver Node, and we instantiate one Network Emulator Node per cellular path. Each emulator runs as a Linux middlebox between the vehicle-side and video-receiver subnets. On each emulator, a traffic-emulation program reads a time-series scenario containing end-to-end link-performance targets (e.g., throughput, delay, and loss) collected using \texttt{iperf2}\footnote{\url{https://sourceforge.net/projects/iperf2/}} during a drive. It periodically updates \texttt{tc}\footnote{\url{https://manpages.ubuntu.com/manpages/xenial/man8/tc.8.html}} on the outgoing interfaces according to the scenario timeline, shaping throughput and adding delay, jitter, and loss to reproduce the observed time-varying user-plane performance along the route. This enables controlled, repeatable comparisons of transport and application configurations under identical replayed conditions\footnote{While downlink bandwidth is workload-dependent, prior work has demonstrated that uplink performance is much less sensitive to the workload~\cite{sentosa2025cellreplay}. Therefore, to emulate end-to-end user-plane throughput and delay dynamics, we use traces measured using \texttt{iperf2} under a fixed workload.}\footnote{For simplicity, we assume that all paths in the multipath setup are continuously active (e.g., through periodic keep-alive traffic), and we focus on emulating user-plane performance with \texttt{iperf2} and \texttt{tc}. However, CloudEmu does not reproduce all cellular factors, such as control-plane latency and RAN-side scheduling effects. Modeling these effects is left for future work.}.

\subsubsection{\textbf{Video-Receiver Node}}
\label{sec:receiver-node}

This node emulates the remote operations center that receives video streams and monitors the vehicles. It runs the receiver-side video-uplink stack, including the multipath gateway server, the video reception pipeline, and the monitoring and logging tools used in our field trials. As shown in \figureref{fig:gui-center-side}, the operator can view the received video while replaying the same network traces under different ABR and multipath settings, enabling repeated assessment of operator-side quality of experience (QoE) and of the communication-parameter tuning workflow.

\subsection{Emulation Scenario Creation}
\label{sec:scenario-file}

An emulation scenario includes a timestamp, GNSS position (latitude and longitude), and network performance data such as throughput, estimated one-way delay, jitter, and packet loss.

\subsubsection{Position and Network Traces Alignment}
\label{sec:time-sync}

During data collection, the position traces were recorded at 1\,Hz, whereas the network traces were sampled every 50\,ms. To align the position traces with the 50\,ms network samples, intermediate positions are linearly interpolated, producing a time-aligned scenario at 50\,ms resolution in which each network-condition sample is paired with a corresponding GNSS coordinate\footnote{Although linear interpolation can introduce meter-level position errors during turning maneuvers at intersections, these errors are usually smaller than the spatial correlation distance (12--15\,m) used in standard spatial-consistency channel models for urban cellular environments~\cite{studyOnChannelV19}. Thus, we expect limited impact on the analysis of replayed user-plane performance at the route level.}.

\subsubsection{Delay Correction}
\label{sec:delay-correct}

As introduced in \secref{sec:emulator-node}, we use \texttt{tc} to reproduce network conditions by shaping throughput and adding delay. However, the delay estimates derived from \texttt{iperf2} UDP probing already include queuing induced by the measurement traffic. If we feed these values into \texttt{tc} while also limiting throughput, we double-count queuing and overestimate the emulated delay, especially under heavy load.

To address this issue, we propose a congestion-induced delay correction method. It flags intervals where throughput stays below $B_{\mathrm{th}}$ and delay exceeds $D_{\mathrm{th}}$ for at least $T_{\mathrm{th}}$, then replaces delay and jitter with averages over windows of length $T_{\mathrm{adj}}$ before and after each interval.
This way, the queuing delay is primarily introduced by the throughput limit in \texttt{tc}, while the configured delay reflects the baseline propagation component. As a result, the emulated delay more consistently matches the measured end-to-end behavior, even under congestion.

\begin{figure}[h]
    \begin{center}
        \includegraphics[width=\linewidth]{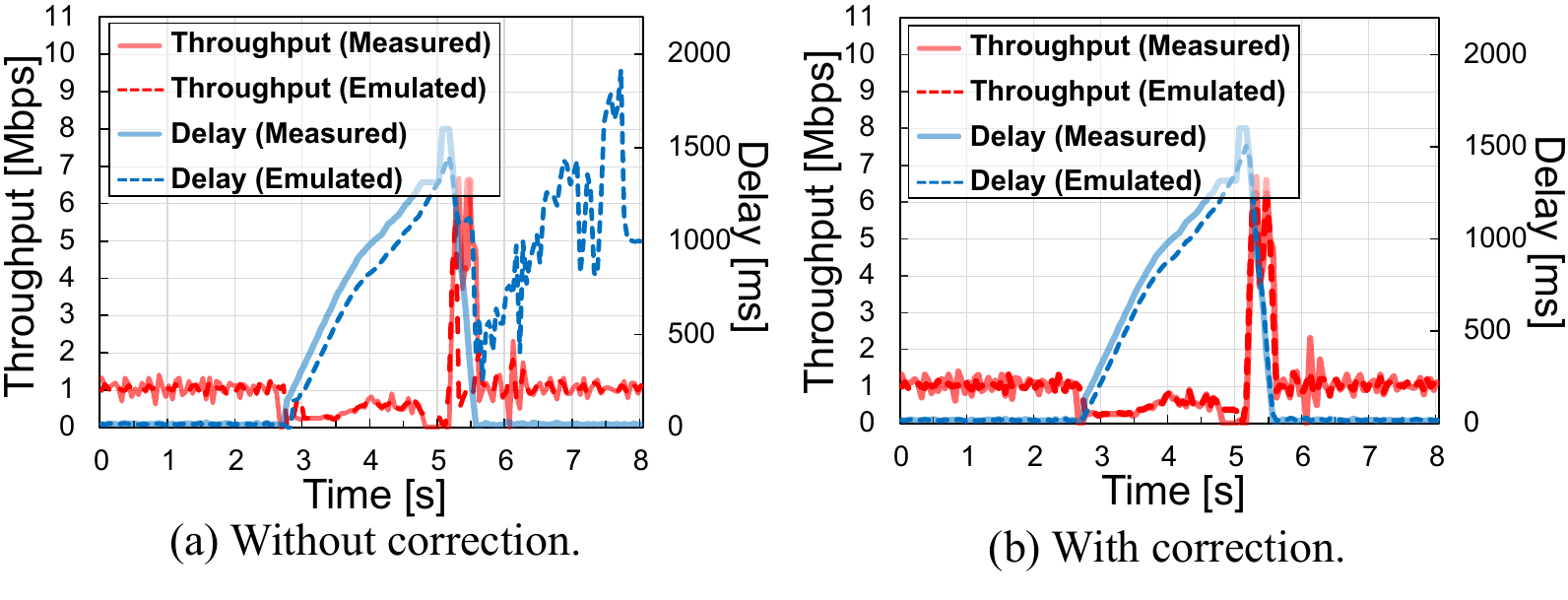}
    \end{center}
    \vspace{-6.5mm}
    \caption{Emulated throughput / delay with and without our correction method.}
    \label{fig:eval}
    \vspace{-3.5mm}
\end{figure}

We evaluated the emulator using a scenario derived from field measurements that contains a representative congestion-induced delay spike. Using UDP probing with \texttt{iperf2} (1\,Mbps offered load and 50\,ms reporting intervals), we compared throughput and estimated one-way delay across (i) the field run, (ii) the emulator driven by the raw scenario, and (iii) the emulator driven by the corrected scenario\footnote{$B_{\mathrm{th}} = 0.7$\,Mbps, $D_{\mathrm{th}} = 50$\,ms, $T_{\mathrm{th}} = 250$\,ms, $T_{\mathrm{adj}} = 1$\,s}. As shown in \figureref{fig:eval}, both scenarios reproduce throughput well, but the raw scenario yields inflated and unstable delay estimates after the congestion episode due to double-counted queuing, whereas the corrected scenario tracks the field delay trace more closely.

%% file: components/03demo.tex
\section{Demonstration Setup}
\label{sec:demo}

\begin{figure}[t]
    \vspace{-1mm}
    \begin{center}
    \subfigure[Route map and vehicle speed.]
    {
        \label{fig:location-with-speed}
        \begin{minipage}{0.51\hsize}
            \centering          
            \includegraphics[width=\linewidth]{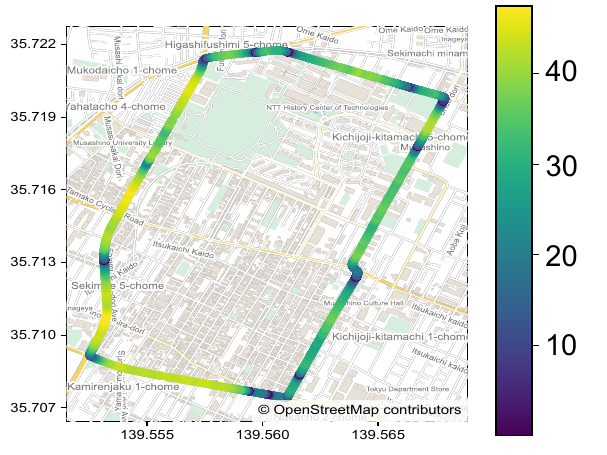}
        \end{minipage}
    }\hspace{-2.3mm}%
    \subfigure[Overview of the demo.]
    {
        \label{fig:demo-machine-setup}
        \begin{minipage}{0.46\hsize}
            \centering          
            \includegraphics[width=\linewidth]{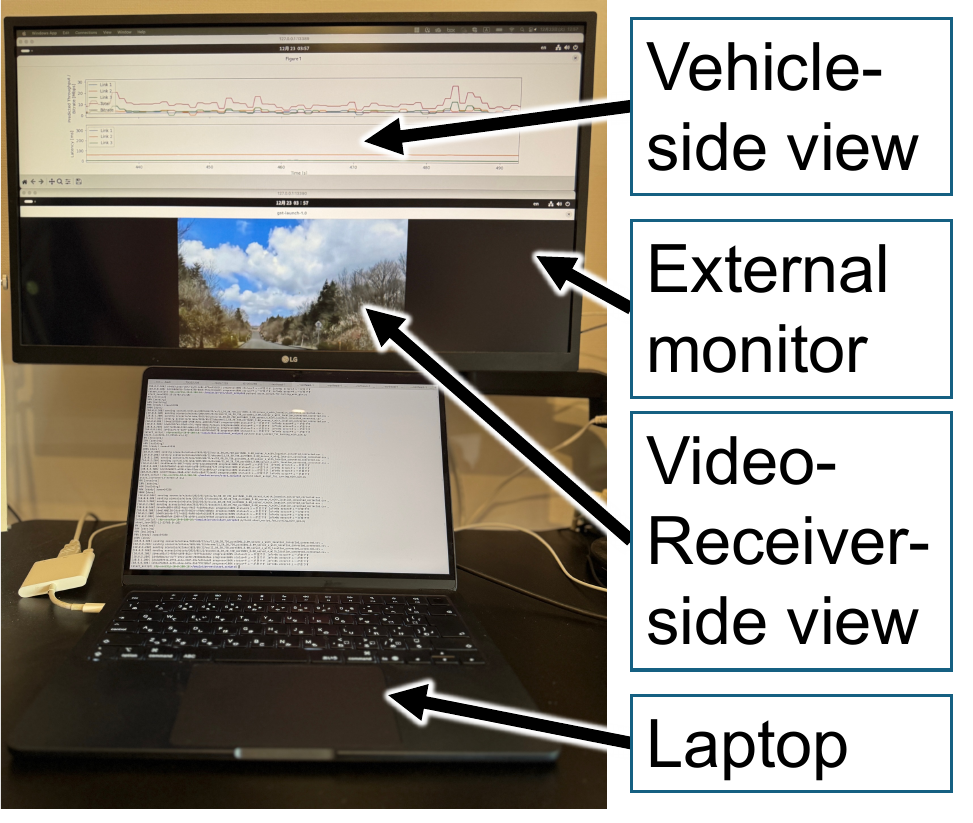}
        \end{minipage}
    }
    \end{center}
    \vspace{-4mm}
    \caption{Overall demonstration setups.}
    \label{fig:setup}
    \vspace{-4mm}
\end{figure}

We demonstrate trace-driven, route-aware emulation of vehicle video uplink communication on CloudEmu. During the demo, we replay cellular and position traces collected along an urban route in Tokyo, as shown in \figureref{fig:location-with-speed}. The cellular trace was collected over three LTE links (from two major mobile operators in Japan) along a 4.8\,km route, and was synchronized with the position trace using timestamps (as discussed in \secref{sec:time-sync}). In the emulation, we stream prerecorded video instead of live in-vehicle camera footage, enabling controlled comparisons under identical replayed trace conditions.

CloudEmu is deployed on Amazon EC2\footnote{\url{https://aws.amazon.com/ec2/}} instances for this demo; during the demo, we use a laptop to access CloudEmu and output both the vehicle-side and video-receiver-side views to an external monitor, as shown in \figureref{fig:demo-machine-setup}.
Attendees can adjust ABR and multipath parameters and then compare video quality and latency under identical replayed trace conditions.
Specifically, attendees can tune (i) ABR thresholds, including bitrate-selection and delay thresholds, (ii) the multipath gateway packet scheduler, and (iii) retransmission frequency, and observe their impact on video uplink streaming under identical replayed trace conditions. Through this tuning, attendees can experience low-cost, reproducible communication-stack testing that would otherwise require on-road trials.